\documentclass[aps,pre,twocolumn,groupedaddress]{revtex4-1}

\usepackage{graphicx}
\usepackage{float}
\usepackage{dcolumn}

\begin{document}

\preprint{}

\title{A Comparison of Genetic Algorithms and Simulated Annealing in Maximizing the Thermal Conductivity of Discrete Massive Chains}


\author{Alexander Kerr}
\author{Kieran Mullen}
\affiliation{Homer~L.~Dodge Department of Physics and Astronomy, The University of Oklahoma, \\ 440 W. Brooks St., Norman, OK 73019}


\date{\today}

\begin{abstract}
	
Functions of chemical composition are complex and discrete in nature making it impossible to optimize them with gradient methods.  Genetic algorithms, which do not use derivative information, are used to maximize the thermal conductivity of one-dimensional classical harmonic oscillators made from a fixed library of  randomly generated molecular units.  The ability for the genetic algorithm to build structures with components having no physical increment is important in optimizing molecules with a library of unrelated polymer units.  The performance of genetic algorithms in this problem is compared with simulated annealing.  Hyper-parameters for these routines are selected from a grid search in order to optimize their expected solution strength.  The solutions found via the genetic algorithm consistently outperform those of simulated annealing at the cost of longer computer time.  Together, these algorithms are able to find thermal conductor candidates that mirror solutions in continuous media.

\end{abstract}


\maketitle

\section{Introduction}\label{sec:intro}

Optimization is a common task in science, engineering and especially material studies in which (un)desirable physical properties are (min)maximized for their implementation in technological devices.  Optimizing molecular structures in particular provides a number of challenges.  Some difficulties originate from the strong coupling between molecules and their environment which greatly affects their physical properties and effectiveness. The evaluation of their properties is also more difficult to calculate and interpret than the bulk properties of macroscopic materials.  Since molecular composition is discrete it is impossible to use methods like gradient descent because there is no way to infinitesimally change chemical composition.  Second, the state space is exponentially large due to the many permutations of molecular units to be considered.  It is impossible to exhaustively search through every possible molecular configuration.  

Despite these challenges, many optimization routines have been studied in material science such as simulated annealing, differential evolution, and particle swarm analysis \cite{inclan}.  Another such optimization process is the genetic algorithm (GA) which uses mechanisms borrowed from Darwinian evolution to naturally select fit solutions to problems \cite{falkenauer}.  GAs, beyond having useful features like their capacity to perform on parallel architectures, have other abilities making them ideal for polymer design including that they: (1) can explore unknown, complex solution surfaces, (2) handle many parameters at once,   (3) return populations of candidates rather than single solutions, and (4) do not require derivative information allowing them to explore discrete spaces (crucial in optimizing composition) \cite{haupt}.  Some material designs have already been served using GAs such as polymer dielectrics \cite{mannodi}, gold nanocatalysts \cite{sen}, among others \cite{0809.1613, 1707.02949}.

We are interested in maximizing the effective thermal conductivity of carbon nanotubes (CNTs).  Despite their extremely high intrinsic, longitudinal thermal conductivity listed at 6000 $\text{W/mK}$ \cite{PhysRevLett.84.4613}, CNTs suffer from a severe Kapitza or boundary  resistance making them ineffective at improving thermal conductivity in  nanocomposites. Simply stated this means the characteristic vibrations of the CNT match poorly with those of the surrounding polymer.  Enhancing CNTs with chemical functionalization at their interfaces may match the thermal impedance between the main bodies and their environment.  If the enhanced CNTs would couple well enough with their plastic matrix, one could develop a material of major technological importance that rivals metals in their thermal conductivity with a fraction of their weight and cost.

This problem is ideal for a study of different approaches to molecular optimization; it is one dimensional and metrics of thermal conductivity are relatively simple to calculate.  For a given library of $N$ molecular units, the number of side chains of length $L$ units is $N^L$, which rapidly exceeds exhaustive computational searches when there is no way to apply continuum optimization techniques to the problem.  In this work we consider a toy model of `1-dimensional' CNTs with singular chains attached to each end.  We test the ability of two particular numerical routines to optimize the composition of these chains, namely simulated annealing (SA) and the GA, and compare their performance.  This comparison is difficult because of their difference in design and considering their vast sets of possible hyperparameters.  This leads to a meta-optimization problem which is the process of finding a set of hyperparameters that give an algorithm the best expected performance for a given problem.  Below we examine certain performance metrics of SA and the GA including their runtime and solution quality.

Section \ref{sec:problem} will detail the scientific aspects of the problem at hand: the thermal conductivity calculation, the 1D CNT systems to be studied, and their numerical representation.  Section \ref{sec:algorithms} will outline the computational techniques of SA and the GA themselves.  These routines will be applied to the thermal conductivity problem in Section \ref{sec:results} where standard parameter sets will be chosen.

\section{Statement of Problem}\label{sec:problem}

In the three sections below we briefly describe the molecular system we are attempting to optimize, the method for calculating the thermal conductivity of a given system, and finally the representation of system in the GA.  

\subsection{1D Molecule}

\begin{figure}[t]
 	\includegraphics[width=3in]{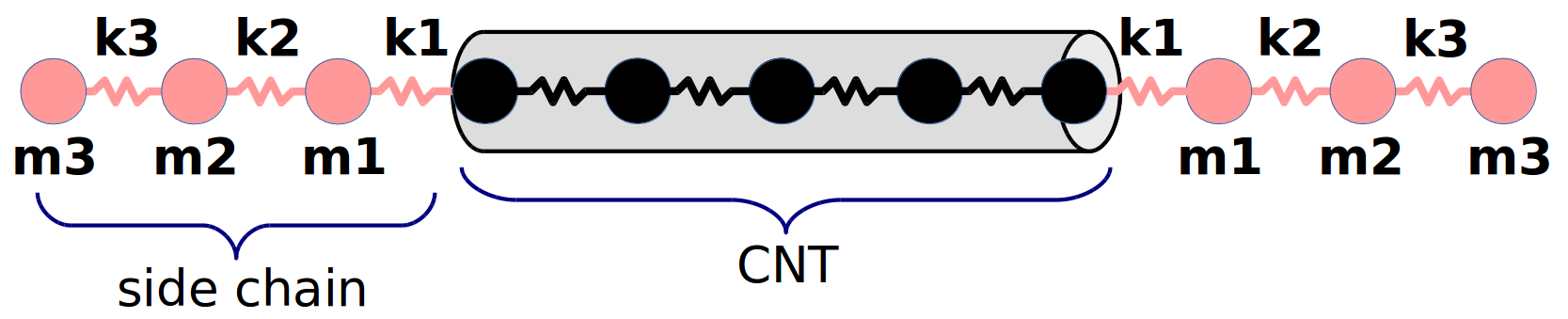}
 	\caption{Schematic representation of a symmetric 1D CNT with end functionalizations.  The middle region is composed of static masses and springs (in black) while the side chains (in red) correspond to parameters $ k_i $, $ m_i $.  Only the end masses ($ m_3 $) would be driven by white noise in the Green's function calculation.\label{fig:cnt-schematic}}
 \end{figure}

  We consider a finite 1-dimensional chain of masses connected by simple harmonic springs.  Our goal is to construct a chain of length $N$ from a library of fixed but random components that will maximize the thermal conductivity across the body of the chain.  The chain in question is divided into three non-overlapping spatial regions: the middle region is composed of intermediate masses and rigid springs, both remaining constant.  The outer two regions are composed of variable masses and springs with each component occupying an allowed, discrete value.  The chain is a mapping of a real molecular system, comprised of a CNT with end functionalizations, to a single dimension.  To reduce the degrees of freedom in the calculation, only chains having inversion symmetry will be evaluated.  Stated differently, only a single side chain is varied but one is bonded to each side of the 1D CNT such that they are mirror images.  The driven masses detailed in Section \ref{subsec:tc} will be limited to the two ends.  Figure \ref{fig:cnt-schematic} displays a schematic representation of such a 1D CNT system.

\begin{figure}[h]
	\includegraphics[width=3in]{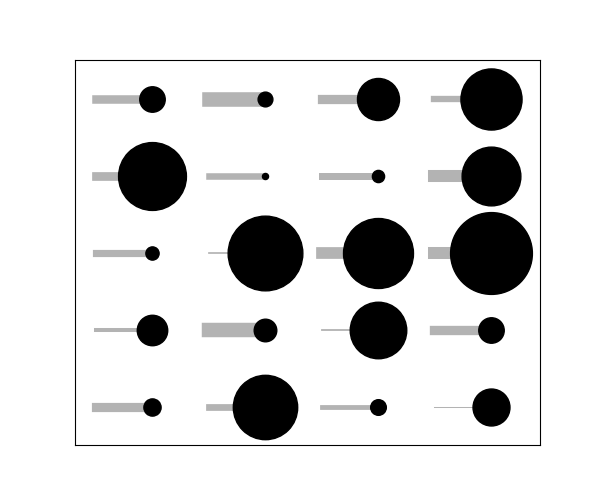}
	\caption{Schematic diagram of one randomly generated library of molecular units from which solutions were built in this experiment.  The width of the line is proportional to the stiffness of the spring while the area of the sphere is proportional to the mass of the unit.   Some of these mass and spring pairings are found in the candidates of Figure \ref{fig:can}.  \label{fig:library}}
\end{figure}

The simplest version of the 1D problem is the one in which every mass and spring of the side chain is independently varied.  Here the expected optimal configuration is the uniform one as this would minimize the vibrational reflections through the system.  If the allowed values are finely spaced, the continuum theory may be used to find a solution \cite{gunawardana}.  However, more interesting developments may be found by adding features such as parameter constraints.  One could fix particular mass and spring values, such as the ends of the chain.  In this paper we will not manipulate individual masses and springs, rather we rely on a set of randomly generated pairs of masses and springs that will act as the molecular units in the problem such as those shown in   The ability for these algorithms to optimize structures with random units would prove valuable in solving problems in which there is no physical or continuous relation between neighboring states.  It is important to note atom types in more realistic calculations \cite{amber} depend on their total chemical environment which could depend on chemical information that is many bonded neighbors away \cite{antechamber}.

\subsection{Thermal Conductivity Calculation}\label{subsec:tc}

We  calculate the thermal conductivity of molecules through the Green's function method detailed in \cite{aitmoussa}.  In such an approach, the molecules under study are relaxed to equilibrium and their effective spring constants between the atoms are determined.  These are used to calculate the normal modes and then the Green's function for interacting damped harmonic oscillators.  The calculation provides an analytical, steady-state solution for the thermal conductivity of interacting masses connected to white noise heat baths.  This technique is an alternative to simulation methods \cite{PhysRevB.71.085417, SALAWAY2014954} which come with a higher computational cost.  The Green's function method comes with a loss in accuracy in general.  In the case where atomic interactions are strictly harmonic the calculation is exact, otherwise  non-linearities are ignored. Such non-linearities are important to obtain precise agreement with experiment, but are not necessary to address the primary bottlenecks in the conduction of heat.  Since this work focusses on the efficiency of GA optimization, we made all interactions harmonic from the outset.

The treatment of a molecule's matrix is also inexact. The heat baths represent the only external forces to the molecule.  In this model the molecule lives in a vacuum save for a few masses which are designated to interact with the baths.  It is between these baths that the heat flows through the system. Again, if the boundary resistance is large, neglecting the interaction with matrix is reasonable.   The effective thermal conductivity is determined from the power driven by the interactions through the cross-section of the CNT.  This power is linearly dependent on the temperature difference of the heat baths.  The calculated quantity is the corresponding linear coefficient
\begin{equation}
\kappa \propto \frac{P}{k_B \Delta T}
\end{equation}
where $ P$ is the steady-state power (heat flux) transferred between the heat baths.  It should be noted that $ \kappa $ in this calculation is not the physical thermal conductivity in $\rm W/m^2 K$   but rather a quantity that has dimensions of inverse time.  Because of these series of approximations, the Green's function method will not be used for accurate evaluations of the properties of molecules.  Instead, the Green's function method will be used to find trends that open the major bottlenecks to thermal conductivity and to select candidate materials.  If improvements are found in the linear contributions to thermal conductivity it is reasonable to expect an overall improvement in some molecules of interest.

\subsection{System Representation}

Simulated annealing and the genetic algorithm handle their solution candidates in the form of numerical strings, or single dimensional arrays of numbers.  In the context of GAs these strings are also called `chromosomes'.  The representation of data points is crucial for the success of numerical applications such as in machine learning in which ideal representations must obey properties such as uniqueness and invariance \cite{1704.06439}.  Fortunately, the 1D systems under study in this manuscript lend themselves to a representation that is unique, information complete, and intuitive.  This representation is a literal data string where a sequence of numbers corresponds exactly to a sequence of masses and springs in the side chains.  Each molecular unit (mass, spring, or pair) is assigned an index in the chromosome and its element maps to its respective physical value in the 1D chain.  The masses and spring strengths in this experiment will have units of $m$ and $ k $ respectively.  In the case of the optimization problem with randomly generated couplings with side chains composed of $N$ masses and symmetry under parity is enforced, the chromosomes have length $N$.

\begin{figure}[t]
	\includegraphics[width=3.5in]{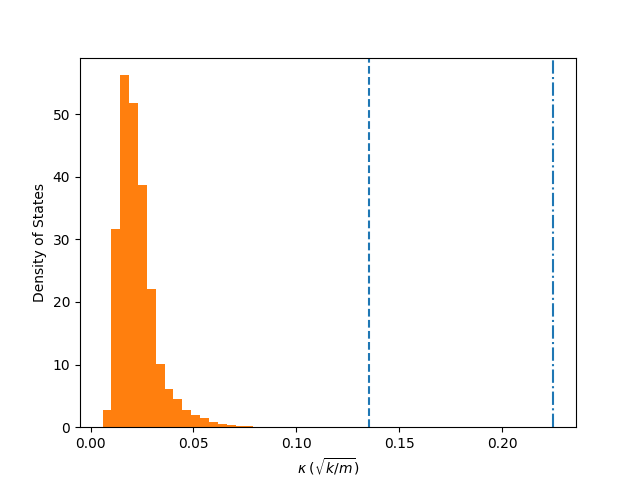}
	\caption{The distribution of thermal conductivities $n(\kappa)$  of $\rm 10^6$ random  1D molecular chains  which have  $ 20^{10} $ possibilities.  The dashed line in the middle of the plot shows  the maximum thermal conductivity found in the  random sample, while the dashed dotted line at the far right is a GA result. \label{fig:tc-dist}}
\end{figure}

To get an intuition of the state-space being investigated, many chains are sampled.  The thermal conductivity of a million random chains in a system with 10 variable units (the generated couplings with 20 possible values for each ($ 20^{10} $ states possible) are calculated.  While the average $ \kappa $ is found to be 0.0227 $ \sqrt{k/m} $, the maximum in the sample is 0.135 $ \sqrt{k/m} $.  The distribution of these evaluations is plotted in Figure \ref{fig:tc-dist} along with a GA result, suggesting the GA is an effective method in selecting molecular chains.

\section{Algorithms}\label{sec:algorithms}

We will use two different meta-heuristics, simulated annealing and the genetic algorithm, to explore the state space of the problem.  Below we describe each method before comparing their efficacy.

\subsection{Simulated Annealing (SA)}

Consider a liquid that is cooled until it solidifies.  If the substance is cooled slowly, the constituent atoms will align themselves to a crystalline state with low energy.  If the liquid is quenched it will end in an amorphous state.  The crystalline structure represents the global energy minimum of the material with respect to the atomic positions. In an attempt to replicate nature's ability to arrive at the energy minimum of thermodynamic systems, an optimization routine was developed known as simulated annealing.  Simulated annealing uses steps from the Metropolis-Hastings algorithm to pick new configurations of an input space.  Instead of using a fixed temperature $T$, this process uses $T$ as a parameter that starts high and is gradually lowered with the hopes of finding the global minimum, akin to liquids becoming crystalline at slow cooling.  Starting with high $T$ allows the algorithm to explore configuration space more freely, making it more likely to find a path to the true global minimum.  To perform simulated annealing, the algorithm needs a 1) set of allowed neighbor moves and 2) a temperature schedule.  The routine is very similar to that of Metropolis-Hastings in that it usually begins with a random initial state $s_0$, then through a number of iterations a set of neighboring states are accepted or rejected according to their transition probabilities.  In this work we define neighboring states as those with a single element changed;  other Monte Carlo moves may yield different efficiencies.  
The major difference between simulated annealing and Metropolis-Hastings is the presence of the temperature schedule which specifies the starting temperature $T_{0}$ and determines the rate in which this parameter decreases, finally reaching $T = 0$ unless other stopping criteria are met.

SA will be applied to problems outlined in Section \ref{sec:problem} using different parameters: namely the annealing schedule and neighboring state definition.  Because $\kappa$ is maximized in this problem while the SA library attempts to \textit{minimize}, it is understood that the negative thermal conductivity is as evaluated as the energy.  The parameter $T$ is understood to be in the same units as the objective function.  The algorithm is given as \cite{otten}:

\begin{center}
	The Simulated Annealing Algorithm
\end{center}
\begin{enumerate}
	\item Begin with an initial state $s_0$ and temperature $T_0$ of the system.  Calculate $E_0$.
	\item For each iteration $i$:
	\begin{enumerate}
		\item Generate a new state $s_{i}$ that neighbors $s_{i-1}$ through a specified MC move.
		\item Calculate the energy of the new configuration $E_{n+1}$.
		\item Define $ a  = \text{min}[1, \exp (-\Delta E / T)]$
		\item Randomly generate a number $r$ uniformly distributed between $0$ and $1$.
		\item Compare $r$ to $a$:
		\begin{enumerate}
			\item If $r \leq a$, accept the new configuration.
			\item Else, reject the new configuration and accept the old one.
		\end{enumerate}
		\item Decrease $T$ according to the temperature schedule.
	\end{enumerate}
	\item Return the final state $s_n$
\end{enumerate}
If successful, the final state $s_n$ is a near energy minimum state of the system.

\subsection{Genetic Algorithm (GA)}

The genetic algorithm is a search method based on the principles of evolutionary biology with the goal of optimizing a supplied fitness function.  The GA allows a population of inputs to evolve over time to maximize this fitness function.  Therefore the fitness corresponds to how well an individual performs; inputs with the highest fitness output are desired.  Inputs will be given as arrays of numbers known as chromosomes.  These chromosomes live in a population comprising of a number of individuals defined by the population size ($ n_{\text{pop}} $)  Each index of a chromosome array corresponds to some parameter interpreted by the fitness function.  

When the algorithm begins, the initial population is filled by individuals comprised of randomized elements.  This population evolves through a selected number of iterations ($n_{\text{epoch}}$) with a set of rules usually determined by the user.  These rules determine the selection of individuals to act as parents of the next generation, the nature of how these individuals mate, and how resulting children mutate.  Parent chromosomes are selected in a manner that emphasizes the most fit individuals.  The first selection method choice is weighted random pairing in which chromosomes are selected at random, weighted by their fitness.  Tournament selection is an alternative method where chromosomes go head-to-head until champions are found to parent the next generation.  

\begin{figure}[t]
	\includegraphics[width=3in]{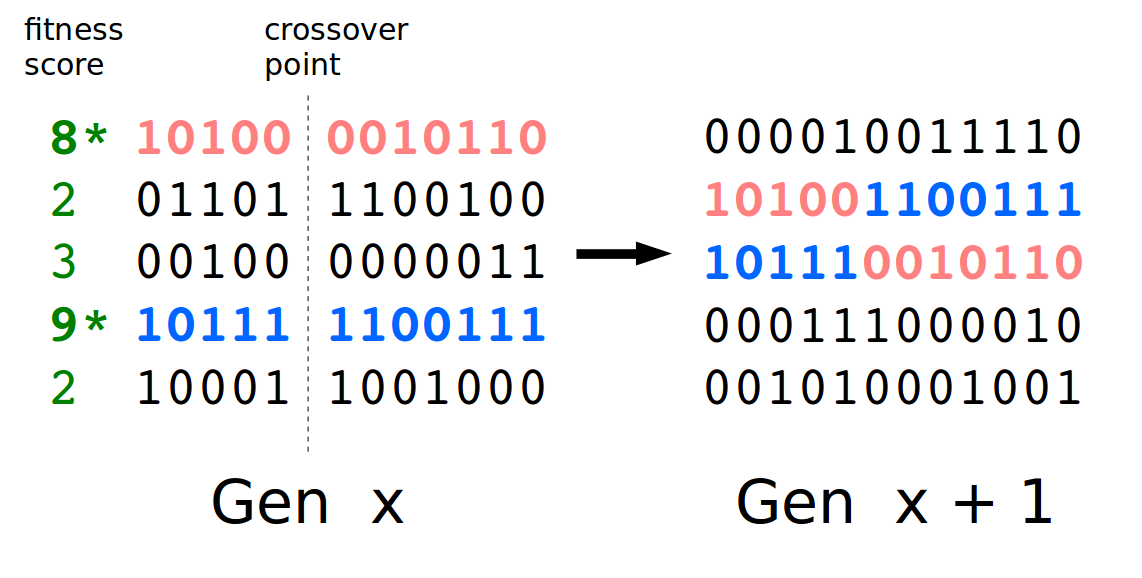}
	\caption{Pictured is a population evolving through one generation in an undescribed problem.  Two chromosomes are selected for breeding based on their fitness.  Their children are formed by gluing their substrings together in a process called single-point crossover.  \label{fig:crossover}  Mutation would occur right after crossover.}
\end{figure}

Once parents are selected, mating is performed via genetic crossover.  During genetic crossover a crossing point is randomly chosen for each parent pairing between $1$ and $L-1$ for chromosomes of length $L$.  A split occurs at this point separating $2$ parts of the resulting children: substring from parent $A$ and substring from parent $B$.  Two children are produced from each parent pairing (conserving the population size) such that the first child inherits parent $A$'s substring before the crossing point, and parent $B$'s substring \textit{after} the crossing point.  The reverse is true for the second child.  Genetic crossover is represented schematically in Figure \ref{fig:crossover}.  After children have been generated they are randomly mutated.  Each element, or gene, of every chromosome is mutated with a probability defined by the mutation rate $ r_{\text{m}} $.  A finite mutation rate supplies ergodicity \cite{vose}. When mutated, the new gene is changed to a random allowed value.  One more GA rule is usually enforced known as elitism.  During elitism, the best chromosomes of one generation are forced into its successor.  Formally stated, the $n_{\text{elite}}$ best chromosomes of generation $n$ are placed into generation $n+1$ after the mutation process.  This guarantees that solutions are, at worst, constant over time and only improve otherwise.

\section{Results}\label{sec:results}

\subsection{Hyperparameter Selection}

\subsubsection{GA vs. GA}

\begin{figure}[t]
	\includegraphics[width=3in]{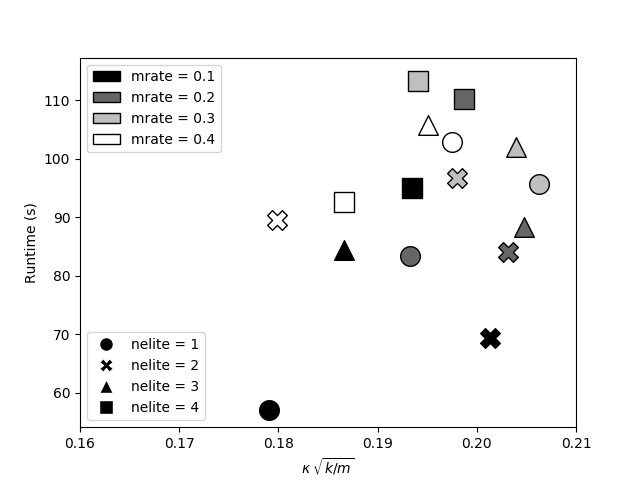}
	\caption{ The distribution of average GA results for different sets of non-zero mutation rate and n-elite.  Some pairs strictly outperform others, such as a mutation rate of 0.2 and n-elite of 1.  
	  \label{fig:rt-tc}}
\end{figure}

There is no way of knowing \textit{a priori} which hyperparameters will secure an optimal solution search.  In this section we implement a simple grid search for the parameters that allow the GA to optimize molecular chain thermal conductivity at a minimal runtime cost.  There is a natural positive correlation between running time and solution strength related to the volume of phase space (ie number of chromosomes) explored.  However we would like to find the best pairing of $ r_{\text{m}} $ and $ n_{\text{elite}} $ with every other parameter being equal, resulting in a 2-dimensional space being lightly sampled.  These two parameters offer the most variability in the calculation when compared to, for example,  $ n_{\text{pop}} $.  The results of the grid search are featured in Figure \ref{fig:rt-tc}.  A mutation rate of 0.3 and $ n_{\text{elite}} = 1 $ appears to strictly outperform other hyperparameter pairs.  The GA experiments are performed with fixed population sizes of 24 and a stopping condition that ends the program when the elite units do not change in 100 generations.  The maximum number of GA generations is 1000.

\subsubsection{SA vs SA}

The SA hyperparameters are: the starting temperature $ T_0 $, annealing rate $ \alpha $ (characteristic decay of the exponential annealing schedule), and the neighbor definition in the Metropolis steps.  In the problem of randomly generated units, there is no physical increment to inform a local neighbor definition.  The choice becomes the number (or fraction) of random genes that get adjusted to random allowed values. One may change a single element of the state vector, every element, or anything in between.  A grid search is performed here as in the preceding section and the following values are selected: $ T_0 = 2000$, $ \alpha = 0.9 $, and a neighbor definition in which a single gene is randomized from the original state.  These parameters were chosen based on the perceived optimal trade-off between runtime and solution strength.  It should be noted that this optimal $ \alpha $ is very low as this parameter is typically very close to unity.  The result is an extreme drop-off in temperature over time that would presumably hinder the SA's ability to sample the space fairly.  Nonetheless, this $ \alpha $ generates the most consistent SA results.

\subsection{GA vs SA}

\begin{figure}[t]
	\includegraphics[width=3.5in]{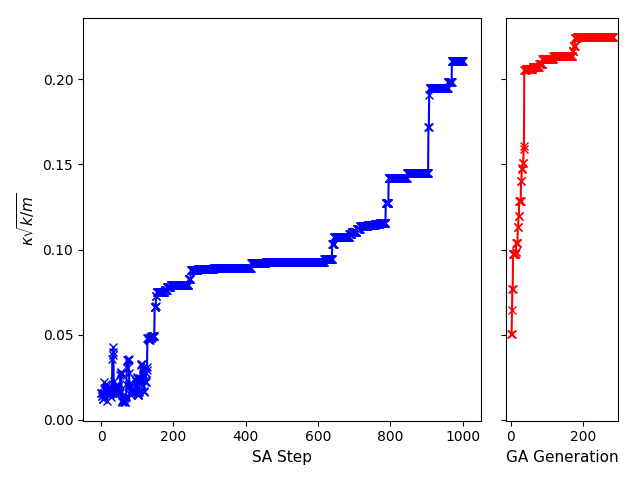}
	\caption{A typical evolution path for the solutions generated by SA and the GA respectively.  While the generation count is low for the GA, each of its iterations is considerably more expensive than a simulated annealing step.\label{fig:evo}}
\end{figure}

\begin{figure}[t]
	\includegraphics[width=3.in]{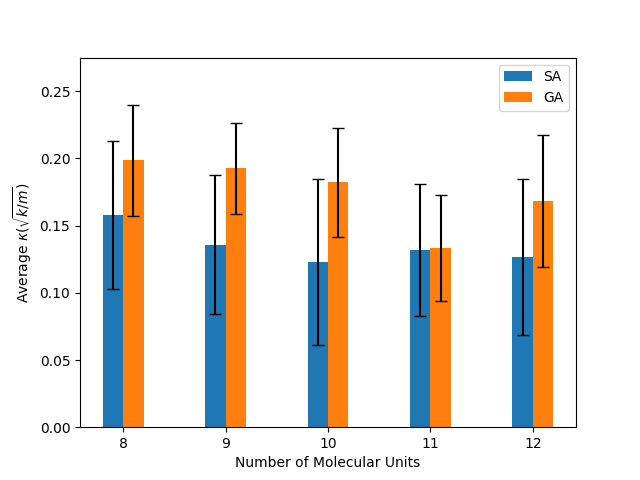}
	\caption{Average $ \kappa $ results for the GA and SA as a function of variable system size.  The error bars signify the standard deviation of the optimization results, showing there are significant fluctuations in the results, even for the GA.  The uncertainties of the GA results are smaller across the board, further contributing to the argument about the GA being a more reliable optimizer \label{fig:gasa-bar}. }
\end{figure}

With their hyperparameters chosen, the GA and SA go head to head.  A typical evolution of these meta-heuristics is featured in Figure \ref{fig:evo}.  Average results of the GA and SA are shown in Figure \ref{fig:gasa-bar}.  The genetic algorithm regularly outperforms simulated annealing in regards to their final solution strengths.  One can depend on the GA to return a population of thermally conductive chains for every experiment.  The cost of this reliability is an increased running time (roughly by factors of up to 5). There is a notable trend of the thermal conductivity decreasing as a function of the side chain length, due to the increased vibrational reflections as mixed molecular units are added.

Not shown are the two cases of  zero mutation rate and  zero elitism.   A zero mutation rate limits the system to simply rearranging the  units of the starting chromosome without changing the type or number.   An elitism of zero selects the best result of the final generation since all previous generations are ignored.

\begin{figure}[t]
	\includegraphics[width=3in]{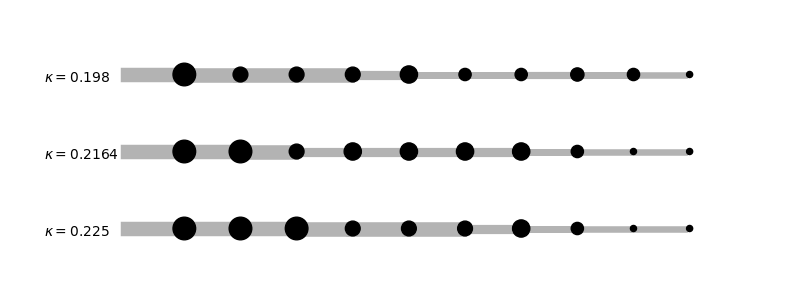}
	\caption{The mass and spring distributions of candidate solutions generated via optimization runs.  The grey bars correspond to springs and their thickness corresponds to their stiffness.  The black circles have an area proportional to the unit's  masses.  These solutions are combinations of the randomly generated mass and spring pairings.  The library of possible molecular units is featured in Figure \ref{fig:library}.  The leftmost spring component is connected to the 1D CNT and the first variable mass, and so on.  The spring constants and masses gradually decrease away from the CNT. \label{fig:can}}
\end{figure}

Three candidate solutions to the problem determined from these algorithms are displayed in Figure \ref{fig:can}.  These candidates have the same striking feature: their springs are very stiff near the CNT interfaces and gradually decrease along the length of the chain.  Their masses also start high and end low at the location of the heat bath interactions.  It should be noted the 1D CNT in each problem was composed of 7 masses at 10 $ m $ each while the springs between them had a uniform strength of 25 $ k $.  The variable spring at the interface exceeds a strength of 20, suggesting the GA correctly matches the impedance of the chains and the CNT.  The speed of sound is decreased at the ends of the chain where the power is driven into the system.  The optimization routines have found chains that gradually increase their speed of sound rather than those that have a severe impedence mismatch at the interface.  These solutions mirror the continuum result which states that a gradual, linear change in the speed of sound maximizes the matching between continuous media.

\section{Conclusion}

Two methods, a genetic algorithm and simulated annealing, were implemented to obtain thermally conductive 1-dimensional chains.  The objective function was evaluated using the Green's function method outlined in \cite{aitmoussa}.  It was shown that sets of hyperparameters can be selected after testing the performance of the algorithm(s) on a grid.  After performing this grid search, it was demonstrated that GA is very effective within this problem scope in finding solutions with an order of magnitude improvement over random sampling.  The effectiveness of the genetic algorithm and simulated annealing were compared and the genetic algorithm repeatedly beats simulated annealing at the cost of longer computer time.  The GA regularly finds chains that match the thermal impedance between the 3 sections.

Concretely we find that optimal elitism can be as small as one chromosome in each generation, and that including more that two does not in general greatly improve results.  Mutations rates of 30-40\% produced the best results in the shortest run time, with lower mutation rates not investigating a large enough state space and larger rates damaging the optimal configurations at such a rate that the efficiency was lowered.  Increasing the number of members in a breeding population simply increased the number of states investigated, and was no better than simply running longer.

These meta-heuristics may be attempted on more physical problem: optimizing the thermal conductivity of 3D functionalized carbon nanotubes.  This 3D problem will have to be parameterized differently; the functionalizations must be be composed of linearly repeating chemically stable polymer units (compare this to the 1D toy model in which sometimes individual masses and springs were tuned).  The interaction strengths of the resulting atomic composition will be determined through a method like that which is outlined in reference \cite{antechamber}.  The interactions depend on individual atomtype assignmments which are sometimes determined by neighboring atoms several sites away.  Higher order bond forces involve atoms across multiple polymer units, making the objective function surface ever more complex.  These points necessitate the clever optimization routines discussed in this manuscript.  There are other sophisticated algorithms and features to the previously discussed procedures that may be considered in future tasks.  One such feature is the generalization of mating in the genetic algorithm to include more than 2 parents in the mating process.  This has been shown to affect the convergence of the GA.

The GA and SA codes in this manuscript were written in python with the \texttt{numpy} package on a Dell XPS 13.  The calculations were entirely serialized but there are natural places to parallelize the code: populations in the GA can be operated on via parallel processes particularly in tournament selection.  One could run several simulated annealings concurrently as they operator independently from each other.

In conclusion we have developed a genetic algorithm to maximize the thermal conductivity of a toy molecular chain.  This algorithm can efficiently find high thermal conductivity candidates in a vast pool.  By examining the winning candidates we can gain insight into the principles guiding their design.

\begin{acknowledgments}
This work  was supported by NSF grant ~\mbox{ DMR-1310407}.
\end{acknowledgments}

\bibliography{references}

\end{document}